\documentclass{iopart}

\usepackage[]{graphicx}

\begin{document}
\jl{1}

\letter {Functional--integral approach to Coulomb fluids\\
in the strong coupling limit}

\author {Hiroshi Frusawa\footnote[1]{e-mail:
frusawa.hiroshi@kochi-tech.ac.jp}}

\address{Soft Matter Laboratory, Kochi University of Technology, Tosa-Yamada, Kochi 782-8502, Japan}

\begin{abstract}
We have developed a field theory for strongly coupled Coulomb fluids, via introducing new functional--integral transformation of the electrostatic interaction energy. Our formalism not only reproduces the Lieb--Narnhofer lower bound, but also bridges logical gaps which previous approaches have involved.
\end{abstract}

\section{Introduction}

Theoretically, there have been thorough studies on strongly coupled plasmas; more than two decades have been passed since the extensive reviews \cite{review}. As a consequence, simulation results for the one component plasma (OCP) have been reproduced precisely by various methods [1--5]. Turning our attention to the two component plasma (or the restricted primitive model electrolytes), however, even the main term of internal energy given by liquid state theories has not coincided with that of crystalline structure \cite{Totsuji-tcp, Fisher1}.

While the fundamental discrepancy has not been resolved, recent simulations and models in the field of soft matter physics have required to investigate more complex Coulomb fluids in the strong coupling regime \cite{CC-review, counter-review}. One of them are colloidal suspensions modeled as either Yukawa fluids \cite{simulations, Rosen-yukawa} or asymmetric two component plasmas, i.e. electrolytes with large asymmetry of size and charge \cite{Fisher2}. Also Monte Carlo simulations have reported that the distribution of strongly coupled counterions dissociated from a macroion is quite different not only from the Poisson-Boltzmann solution, but also from that of the 2-dimensional OCP formed on the macroion surface due to the freedom of extra 1-dimension \cite{counter-review, Moreira}.

The necessity for addressing the advanced issues is prompting one to explore strong coupling theories more systematic and general [14--17]. In our previous work \cite{Frusawa}, a new field-theoretic formulation has thus been devoted to explaining the above counterion electrostatics. This letter will now apply to the OCP, the well--established system, our formalism with considerable improvements, and will demonstrate the relevance through revisiting the lower energy bound of the OCP in the strong coupling limit (SCL).

\section{Lieb--Narnhofer lower bound revisited}

{\itshape Rescaling the model system}--- Let us consider the OCP which consists of $N$ particles with electric charge $Ze$ embedded in a neutralizing background of its volume $\widetilde{L}^3$. As is well known, the OCP is characterized by the Coulomb-coupling constant $\Gamma=Z^2e^2/(4\pi\epsilon k_BT\,\widetilde{a})$ and the Coulomb interaction with large coupling constant ($\Gamma>>1$) has been referred to as "strong coupling", where $\epsilon$ is the dielectric permittivity, $k_BT$ the thermal energy, and $\widetilde{a}$ the Wigner-Seitz (WS) radius defined by $(4\pi\widetilde{a}^3/3)N=\tilde{L}^3$. 

\vspace{4pt}
For clarifying the $\Gamma$-dependence of the OCP, we will rescale the system by the WS radius $\widetilde{a}$. Here, in order to avoid confusion about symbols, we would like to make it clear that tildes are attached to original values and not to the rescaled ones for abbreviating the notation of the rescaled expressions, and that all of the normalized symbols without tildes are dimensionless. For example, correspondences are the following: scalars (the system size $\widetilde{L}$ and the WS length $\widetilde{a}$) transform to $L=\widetilde{L}/\widetilde{a}$ and $a=\widetilde{a}/\widetilde{a}=1$, and vectors of the particles' positions ${\bf \widetilde{r}}_i$ ($i=1,2,\cdots,N$) and of the separation ${\bf \widetilde{r}}={\bf \widetilde{r}}_i-{\bf \widetilde{r}}_j$, respectively, to ${\bf r}_i=\widetilde{\bf r}_i/\widetilde{a}$ and ${\bf r}=\widetilde{\bf r}/\widetilde{a}$; the differential form is rescaled as $d{\bf r}=d\widetilde{{\bf r}}/\widetilde{a}^3$ while the smeared number density in the rescaled system is given by $\overline{n}=N/L^3=N\widetilde{a}^3/\widetilde{L}^3$, therefore we have the reparametrization invariance $\overline{n}\,d{\bf r}=(N/\widetilde{L}^3)\,d\widetilde{\bf r}$. To be noted, the rescaled form of the smeared concentration, $\overline{n}=N\widetilde{a}^3/\widetilde{L}^3$, is not a variable but is equal to the constant, $\overline{n}=3/(4\pi)$, as found from the above definition of the WS radius $\widetilde{a}$.

\vspace{4pt}
\noindent
{\itshape Ewald-type identity and its inequality condition}--- Denoting the excess internal energy per ion in the $k_BT$-unit by $u\equiv U/(Nk_BT)$, the Ewald hybrid expression $u=(\overline{n}/2)\int\,d{\bf r}\; h({\bf r})
   \left[\,
   \phi({\bf r})-\theta({\bf r})+\theta({\bf r})\,
   \right]$, valid for any auxiliary function $\theta({\bf r})$, reads \cite{Rosen-high, Rosen-lower}
   \begin{eqnarray}
   \fl\qquad
   u=\frac{\overline{n}}{2}\int\,d{\bf r}\; h({\bf r})
   \left[\,
   \phi({\bf r})-\theta({\bf r})\,
   \right]+\frac{1}{2}\int\,\frac{d{\bf k}}{(2\pi)^3}\,S({\bf k})\theta(-{\bf k})-\frac{1}{2}\int\,d{\bf r}\;\delta({\bf r})\,\theta({\bf r}),
   \end{eqnarray}
where the radial distribution function $g({\bf r})$ is replaced by the total correlation function $h({\bf r})=g({\bf r})-1$ considering the electrical neutrality, $\phi({\bf r})=\Gamma/|{\bf r}|$ is Coulomb interaction potential, and the structure factor $S({\bf k})$ is the Fourier transform of $\delta({\bf r})+\overline{n}\,h({\bf r})$. The convexity conditions, $g({\bf r})=1+h({\bf r})\geq 0$ and $S({\bf k})\geq 0$, then lead to
   \begin{eqnarray}
   u\geq
   -\frac{\overline{n}}{2}\int\,d{\bf r}\; 
   \left[\,
   \phi({\bf r})-\theta({\bf r})\,
   \right]-\frac{1}{2}\int\,d{\bf r}\;\delta({\bf r})\,\theta({\bf r})
   \equiv u_L\{\theta\}.
   \label{inequality}
   \end{eqnarray}
The best lower bound has been evaluated from optimizing the above functional $u_L\{\theta\}$ with respect to $\theta({\bf r})$ \cite{Rosen-lower, Lieb}.

\vspace{4pt}
\noindent
{\itshape Onsager's smearing}--- Following Lieb and Narnhofer \cite{Lieb}, let us specify the auxiliary function $\theta({\bf r})$ of the form,
    \begin{equation}
    \theta({\bf r};\,\{q\},\,b)
    \equiv\theta_q({\bf r})=\int d{\bf x}\,d{\bf y}\;
    \phi({\bf r}+{\bf x}-{\bf y})\,
   q({\bf x})\,q({\bf y}),
   \label{def-theta}
    \end{equation}
where ${\bf x}$ and ${\bf y}$ are internal vectors of charged spheres (or Onsager balls) whose charge distribution and radius are $Zeq$ and radius $b\,(\leq a)$ equally, and the integrand $\phi({\bf r}+{\bf x}-{\bf y})q({\bf x})\,q({\bf y})$ represents the Coulomb interaction between a point ${\bf x}$ of one ball and another ${\bf y}$ of the other sphere (see Fig. 1). The specified auxiliary function $\theta_q({\bf r})$ therefore corresponds to the Coulomb interaction potential between Onsager balls. Moreover, since the normalization condition $\int_{|{\bf x}|\leq b} d{\bf x}\;q({\bf x})=1$ is imposed, the auxiliary interaction potential $\theta_q({\bf r})$ between non--overlapping balls is the same as the bare Coulomb interaction: $\theta_q({\bf r};|{\bf r}|\geq 2b)=\phi({\bf r})$. This property of the auxiliary potential implies that the Onsager system coarse-grains the point charges within the range $b$.

\begin{figure}[htbp]
	\begin{center}
	\includegraphics[width=5.5cm]{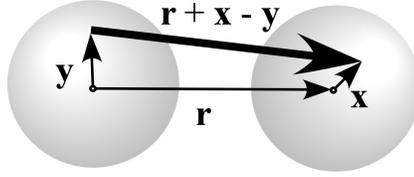}
	\end{center}
	\caption{A configuration of two Onsager balls illustrates the vector, ${\bf r}+{\bf x}-{\bf y}$, explained in the text. In the strong coupling limit, the spheres are in contact with each other because the radius $b$ is found to be equal to the Wigner-Seitz one $a$ \cite{Lieb}.}
\end{figure}

\vspace{4pt}
The minimization conditions with respect to $\theta_q$ (i.e. $\delta u_L/\delta q=\partial u_L/\partial b=0$) then yield the Lieb--Narnhofer lower bound in the strong coupling regime ($\Gamma>>1$): $u_L\{\,\theta_q^{\mathrm {min}}\,\}=-0.9\Gamma$, where the optimized charge distribution is of the forms, $q_{\mathrm{min}}({\bf x})=\Theta(|{\bf x}|-b_{\mathrm {min}})$ and $b_{\mathrm {min}}=a$ \cite{Rosen-lower, Lieb}.

\vspace{4pt}
\noindent
{\itshape Open problems}--- We would like to point out logical leaps which the conventional discussions have made:
  \begin{itemize}
   \item There are no formulations to show that the excess internal energy $u$ given by eq. (1) is reduced to the functional $u_L\{\theta_q\}$ defined by eq. (\ref{inequality}) in the SCL ($\Gamma\rightarrow\infty$).
  \item Since the above framework is based on both the Ewald-type identity and the convexity conditions, it has not been clarified why the trial interaction potential $\theta_q$ (or $q$ and b) should be minimized to know the best lower bound.
  \end{itemize}
These will be addressed in the last section, after deriving the Lieb--Narnhofer bound field-theoretically.

\section{Variational approach}

\noindent
{\itshape Reference system}--- Let us take the reference system constituted of the above Onsager balls. The Helmholtz free energy $F_0$ then reads $\exp\,\left(-F_0\{\theta_q\}\right)
    = \mathrm{Tr}_{\mathrm{cl}}\>\exp
    \left(
    -U_q\{\hat{\rho}\}+U_{q}^{\mathrm{self}}
    \right)$. Here, for brevity, we introduce the classical operator $\mathrm{Tr}_{\mathrm{cl}}=(N!)^{-1}\int d{\bf r}_1\cdots d{\bf r}_N$, set the de Broglie thermal wavelength unity, and represent all energies ($F_0$, $U_q$, $J$, etc.) in the $k_BT$--unit. With the potential $\theta_q$ of the form (\ref{def-theta}), the interaction energies are expressed as follows: $U_{q}\{\hat{\rho}\}= (1/2)\int d{\bf 1}\,d{\bf 2}\;\hat{\rho}({\bf 1})\,\hat{\rho}({\bf 2})\,\theta_q({\bf 1}-{\bf 2})$ and $U_{q}^{\mathrm{self}}=N\theta_q(0)/2$, where we set that $\hat{\rho}({\bf 1})\equiv\sum_{i=1}^N\delta({\bf 1}-{\bf r}_i)-\overline{n}$.

\vspace{4pt}
\noindent
{\itshape Gibbs-Bogoliubov inequality}--- The real system consisting of point charges is recovered from replacing an arbitrary function $q$ by the Dirac delta $\delta$ in eq. (\ref{def-theta}). Denoting the input by the subscript $\delta$, the associated free energy $F$ is expressed as $e^{-F}=\mathrm{Tr}_{\mathrm{cl}}\>\exp\,(-U_{\delta}\{\hat{\rho}\}+U_{\delta}^{\mathrm{self}}\>)$. We aim to reach the true free energy $F$ by exploiting the Gibbs-Bogoliubov inequality \cite{Gibbs},
  \begin{equation}
   F\leq F_0+\left<\,
  (U_{\delta}\{\hat{\rho}\}-U_{\delta}^{\mathrm{self}})
  -(U_q\{\hat{\rho}\}-U_{q}^{\mathrm{self}})
  \,\right>_0
  \equiv F_v,
  \label{GB}
  \end{equation}
where $<\mathcal{O}>_0$ represents the average for the reference system: $<\mathcal{O}>_0=(1/e^{-F_0})\;\mathrm{Tr}_{\mathrm{cl}}\>
  \mathcal{O}\>\exp
    \left(
    -U_q\{\hat{\rho}\}+U_{q}^{\mathrm{self}}
    \right)$. With use of the total correlation function $h_0({\bf r})$ in the reference system, the variational free energy $F_v$ defined in eq. (\ref{GB}) reads
  \begin{equation}
  F_v\,\{\theta_q\}=F_0+\frac{N\overline{n}}{2}
  \int_{|{\bf r}|\leq 2b}d{\bf r}\,
  h_0({\bf r})[\,
  \phi({\bf r})-\theta_q({\bf r})
  \,],
  \label{Fv}
  \end{equation}
where the integration range is specified considering that $\phi({\bf r})-\theta_q({\bf r})=0$ in the region $|{\bf r}|\geq 2b$. Equations (\ref{GB}) and (\ref{Fv}) imply that the reference system is to be selected to minimize the variational free energy $F_v$.


\section{Reference free energy $F_0$ in the SCL}

Manipulating the interaction energy $U_{q}\{\hat{\rho}\}$, the present section reveals what term is negligible in the SCL. The formulations are roundabout at first glance, but relevant and indispensable to taming strongly-coupled Coulomb fluids.

\subsection{Manipulation of the interaction energy, $U_{q}\{\hat{\rho}\}$}

The steps are threefold. First we insert density field $\{\rho\}$ as usual. Next, instead of eliminating the $\rho$--field by the Gaussian--integration, we further introduce a potential field $\{\psi\}$ via Dirac delta functional. Lastly, the Hubbard-Stratonovich transformation of the $\psi$--field adds another density field $\{c\}$.

\vspace{4pt}
\noindent
{\itshape Step 1: Inserting density field $\{\rho\}$}--- Following the standard procedure \cite{Orland}, the first transformation into functional--integrals exploits the identity for the Fourier-transformed delta functional: $1=\int D\rho\,D\varphi\>\exp\left[\,
    i(\rho-\hat{\rho})\cdot\varphi\,
    \right]$, where $f\cdot g\equiv\int d{\bf 1}\,f({\bf 1})\,g({\bf 1})$. Inserting the unity term into $e^{-U_{q}\{\hat{\rho}\}}$, we have $e^{-U_{q}\{\hat{\rho}\}}=\int D\rho\,D\varphi\>\exp\left(
    -U_{q}\{\rho\}
    +i(\rho-\hat{\rho})\cdot\varphi
    \right)$.
    
\vspace{4pt}
\noindent
{\itshape Step 2: Potential field $\{\psi\}$ introduced by hand}--- It is tempting to proceed to Gaussian-integrate over the $\rho$--field because $U_{q}\{\rho\}$ is quadratic. Nevertheless, we would rather add the potential field $\{\psi\}$ than subtract through the following identity:
    \begin{eqnarray}
    1
    =\,\int D\psi\;
    {\rm Det}\left(
    \frac{-\nabla^2}{4\pi\Gamma}
    \right)\>
    \prod_{\{\bf 1\}}
    \>\delta\left[
    \frac{-\nabla^2\psi({\bf 1})}{4\pi\Gamma}
    -\rho({\bf 1})\right]
    \equiv \int D\psi\; \Delta\{\psi, \rho\}.
    \label{delta-psi}
    \end{eqnarray}
The Dirac delta functional defines the potential $\psi$ as $\nabla^2\psi=-4\pi\Gamma\rho$ which is identical to the Poisson equation, $\nabla^2(k_BT/Ze)\widetilde{\psi}({\bf \widetilde{1}})=-Ze\widetilde{\rho}({\bf \widetilde{1}})/\epsilon$, in the original scale with tildes due to the correspondences: $\nabla^2=\widetilde{a}^2\widetilde{\nabla}^2$ and $\rho=\widetilde{a}^3\widetilde{\rho}$. In other words, $\widetilde{\psi}$ is the Coulomb potential in the unit of $k_BT/Ze$. Inserting again the above identity into $e^{-U_q\{\rho\}}$, we have
    \begin{eqnarray}
    e^{-U_{q}\{\rho\}}=\int D\psi\;\>\Delta\{\psi,\rho\}\;
    \exp\left(
    -U_{q}\{\psi\}\,
    \right)\nonumber\\
    U_{q}\{\psi\}=
    \frac{1}{8\pi\Gamma}\int d{\bf 1}\;d{\bf x}\,d{\bf y}\;\;
   \nabla\psi({\bf 1})\cdot\nabla\psi({\bf 1}+{\bf x}-{\bf y})
   \;
   q({\bf x})\,q({\bf y}),
   \label{u-psi}
    \end{eqnarray}
where use has been made of the following relations: $\rho({\bf 1})=-\nabla^2\psi({\bf 1})/(4\pi\Gamma)$, $\int d{\bf 2}\,\phi({\bf 1}+{\bf x}-{\bf 2}-{\bf y})\,\rho({\bf 2})=\psi({\bf 1}+{\bf x}-{\bf y})$, and $(\nabla^2A)\,B=\nabla\cdot(AB)-\nabla A\cdot\nabla B$.

\vspace{4pt}
\noindent
{\itshape Step 3: The Hubbard-Stratonovich transformation}--- Since the form (\ref{u-psi}) of $U_{q}\{\psi\}$ is quadratic, it is possible to perform the Hubbard-Stratonovich transformation as follows:
    \begin{eqnarray}
    \fl\quad
    e^{-U_{q}\{\psi\}}=\frac{1}{\int Dc\;e^{-U_{q}\{\psi\equiv 0,c\}}}
    \>\int Dc\>
    \exp\left(
    -U_{q}\{\psi,c\}\,
    \right)
    \label{e-psi}\\
    \fl\quad
   U_{q}\{\psi, c\}
   =
   \frac{1}{2}\int d{\bf 1}\,d{\bf 2}\;d{\bf x}\,d{\bf y}\;
   \frac{q^{-1}({\bf x})\,q^{-1}({\bf y})}{|{\bf 1}+{\bf x}-{\bf 2}-{\bf y}|}
   \,
   c({\bf 1})\,c({\bf 2})
   +
   \frac{i}{\Gamma^{1/2}}\,
   c\cdot\psi.
   \label{u-psi-c}
   \end{eqnarray}
Only the $\psi$--linear term has the $\Gamma$--dependence proportional to $\Gamma^{-1/2}$, which suggests the possibility of the strong coupling expansion.
   
\vspace{4pt}
\noindent
{\itshape Result: Four-field representation}--- Combining the three steps provides the following four--field expression:
   \begin{eqnarray}
   e^{-U_{q}\{\hat{\rho}\}}&=&
   \frac{1}{\int Dc\;e^{-U_{q}\{\psi\equiv 0,\,c\}}}\>
   \int D\rho\,D\varphi\,D\psi\,Dc\;\;
   \Delta\{\psi,\rho\}\nonumber\\
   &&\quad\qquad\qquad\qquad\qquad\times
   \exp\left[\,
   -U_{q}\{\psi, c\}
    +i(\rho-\hat{\rho})\cdot\varphi\,
   \right]
   \label{four-fields}
   \end{eqnarray}
with $\Delta\{\psi,\rho\}$ and $U_{q}\{\psi, c\}$ defined in eqs. (\ref{delta-psi}) and (\ref{u-psi-c}).

\subsection{Approximate form in the limit $\Gamma\rightarrow\infty$}

We would like to validate that the second term on the right hand side of eq. (\ref{u-psi-c}) is fairy negligible in the SCL ($\Gamma\rightarrow\infty$). To this end, we give the Fourier-transformed expression,
   \begin{eqnarray}
   U_{q}\{\psi, c\}
   =
   \sum_{{\bf k}}\>2\pi\,\left(
   \frac{c_{{\bf k}}c_{-{\bf k}}}{k^2\,q_{\bf k}q_{-{\bf k}}}
   \right)+\frac{i}{\Gamma^{1/2}}\,c_{\bf k}\,\psi_{-{\bf k}},
   \end{eqnarray}
where $|{\bf k}|=\mathrm{k}$, and the denominator $k^2q_kq_{-k}$ of the first term on the right hand side is regarded as the Fourier component of $|\nabla_{{\bf x}}q|^2$. If this denominator increases with larger wavenumber and becomes comparable to $\Gamma^{1/2}$, it is not always justified to ignore the second term proportional to $1/\Gamma^{1/2}$; the approximation holds only in the coarse-grained scale \cite{Frusawa}. Due to the Onsager's smearing of the reference system, however, $k^2\,q_{\bf k}q_{-{\bf k}}$ keeps finite. For example, this can be checked from the relation $\lim_{\mathrm{k}\rightarrow\infty}(\mathrm{k}^2\,q_{\bf k}q_{-{\bf k}})=\lim_{\mathrm{k}\rightarrow\infty}\sigma^2 k^2[\sin (\mathrm{k}b)/(\mathrm{k}b)]^2\leq(\sigma/b)^2$ for a rapid distribution which changes abruptly at the ball surface: $q({\bf x})=\sigma\delta(|{\bf x}|-b)$ and $\sigma=1/4\pi b^2$.

\vspace{4pt}
The above discussions verify that $\lim_{\Gamma\rightarrow\infty}U_{q}\{\psi, c\}= U_{q}\{\psi\equiv 0,\,c\}$. The four-field representation (\ref{four-fields}) is then reduced to the three-field expression which simply yields unity:
    \begin{eqnarray}
   \lim_{\Gamma\rightarrow\infty}
   e^{-U_{q}\{\hat{\rho}\}}=
   \int D\rho\,D\varphi\,D\psi\;\;
   \Delta\{\psi,\rho\}\;
   \exp\left[
  \>i(\rho-\hat{\rho})\cdot\varphi
   \right]=1,
   \end{eqnarray}
where the $\psi$--field integration gives $1=\int D\psi\;\Delta\{\psi, \rho\}$, therefore $\lim_{\Gamma\rightarrow\infty}
   e^{-U_{q}\{\hat{\rho}\}}=
   \int D\rho\,D\varphi\;\;
   \exp\left[
   \,i(\rho-\hat{\rho})\cdot\varphi
   \right]=1$.


\vspace{4pt}
We have thus arrived at the limiting interaction energy, $\lim_{\Gamma\rightarrow\infty}U_{q}\{\hat{\rho}\}= 0$, meaning that violating electrical neutrality is forbidden even locally. In this SCL approximation, the reference free energy $F_0$ takes such a simple form as
     \begin{eqnarray}
     \lim_{\Gamma\rightarrow\infty}
   F_0\{\theta_q\}=
   -\frac{N}{2}\,\theta_q(0)+\int d{\bf r}\;
   \overline{n}\ln \overline{n}-\overline{n},
   \label{ref-free-app}
   \end{eqnarray}
corresponding merely to the mean-field free energy. 

\section{Variational energies in the SCL}

\noindent
To evaluate the perturbative contribution given in eq. (\ref{Fv}), we need to find the density-density correlation between charged balls in the reference system. Since the above section shows that the interactions between Onsager balls are irrelevant in the SCL, we have the limiting behavior $g_0({\bf r})\equiv1+h_0({\bf r})\rightarrow 0$. Equation (\ref{Fv}) hence reads
   \begin{eqnarray}
   \lim_{\Gamma\rightarrow\infty}
   F_v\{\theta_q\}=
   \lim_{\Gamma\rightarrow\infty}
   F_0\{\theta_q\} -\frac{N\overline{n}}{2}\int_{|{\bf r}|\leq 2b}d{\bf r}
   \,[\,
  \phi({\bf r})-\theta_q({\bf r})
  \,],
  \label{Fv-result}
   \end{eqnarray}
where the reference free energy $F_0$ is of the form (\ref{ref-free-app}). Recalling that $\phi$ and $\theta_q$ are proportional to $\Gamma$, the variational internal energy $u_v\equiv\Gamma
   \,(\partial F_v\{\theta_q\}/\partial\Gamma)$ is obtained from eq. (\ref{Fv-result}) as
   \begin{equation}
   \lim_{\Gamma\rightarrow\infty}u_v\{\theta_q\}=u_L\{\theta_q\};
   \label{proof}
   \end{equation}
see eq. (\ref{inequality}).

\vspace{4pt}
The Gibbs-Bogoliubov inequality (\ref{GB}) says that the best free energy $F_v\{\theta_q^{\mathrm{min}}\}$ is obtained from minimizing the above expression (\ref{Fv-result}): $\lim_{\Gamma\rightarrow\infty}
   \delta F_v/\delta q=\partial F_v/\partial b=0$. Moreover, in the SCL, it is the same thing that minimizes $F_v$ and $u_v$ (or $u_L$) with respect to $\theta_q$. Our formalism thus reproduces the Lieb--Narnhofer lower bound: $\lim_{\Gamma\rightarrow\infty} u_v\{\theta_q^{\mathrm{min}}\}=u_L\{\theta_q^{\mathrm{min}}\}=-0.9\Gamma$.

\section{Concluding remarks}

Finally, let us consider the questions posed at the end of section 2, looking back at the arguments we made. Roughly speaking, the proof of eq. (\ref{proof}) has been offered, and the variational approach itself forms the basis of the minimization scheme by Lieb--Narnhofer; it then seems that the missing link described in "{\itshape Open problems}" has been almost found. The supplementary explanations of the following respects, however, remain to be added: ({\itshape S1}) underlying physics of the reference system which selects the mean-field picture in the SCL, and ({\itshape S2}) the connection between the Gibbs-Bogoliubov inequality and the best lower bound of the free energy.

\vspace{4pt}
({\itshape S1}) Inherently, the mean-field theory is the saddle-point approximation valid in the weak coupling regime, $\Gamma <<1$ \cite{Netz}. Some insight is hence required to explain the mathematical result that the reference free energy (\ref{ref-free-app}) is of the same form as the mean-field one in spite of the SCL. We focus on the indistinguishability between the mean-field system smeared overall and the close packing of Onsager's charged  balls. The similarity gives an interpretation that the mean-field picture mimics the frozen system filled with the Onsager balls inside which charges are cancelled by the background; indeed, the fake non-correlation of the reference system in the SCL approximation has led to the vanishing of the radial distribution function, $g_0({\bf r})\rightarrow 0\>(|{\bf r}|\leq 2a)$, which should actually be associated with the non-overlapping of frozen balls.

\vspace{4pt}
({\itshape S2}) Recently it has been proved that the mean-field free energy with repulsive interaction potential is the exact lower bound \cite{mean}, and our limiting reference free energy $\lim_{\Gamma\rightarrow\infty}F_0\{\theta_q\}$, equal to the mean--field one, is just the case: $\lim_{\Gamma\rightarrow\infty}F_0\{\theta_q\}\leq F_0\{\theta_q\}$. Therefore, considering also the inequality $h_0({\bf r})\geq -1$, the limiting variational free energy (\ref{Fv-result}) is found to be the lower bound of $F_v$ give by eq. (\ref{Fv}):
     \begin{eqnarray}
   \lim_{\Gamma\rightarrow\infty}F_v\{\theta_q\}
   =\lim_{\Gamma\rightarrow\infty}F_0\{\theta_q\}
   -\frac{N\overline{n}}{2}\int d{\bf r}\>[\,
  \phi({\bf r})-\theta_q({\bf r})\,]
   \leq
   F_v\{\theta_q\},
   \label{var-inq}
   \end{eqnarray}
which is valid for any auxiliary function $\theta_q$. In principle, it is then possible for an ideal function $\theta^{\mathrm{id}}_q$ to realize $F_v\{\theta^{\mathrm{id}}_q\}=F$ with an arbitrary coupling constant $\Gamma$; to be noted, however, an ideal function $\theta_q^{\mathrm{id}}$ in the case of finite coupling constant cannot be the best, $\theta_q^{\mathrm{min}}$, for $\Gamma\rightarrow\infty$. The relation (\ref{var-inq}) and the Gibbs-Bogoliubov inequality (\ref{GB}) thus lead to
   \begin{eqnarray}
   \lim_{\Gamma\rightarrow\infty}F\approx
   \lim_{\Gamma\rightarrow\infty}F_v\{\theta_q^{\mathrm{min}}\}\leq
   \lim_{\Gamma\rightarrow\infty}F_v\{\theta_q^{\mathrm{id}}\}\leq
   F_v\{\theta_q^{\mathrm{id}}\}=F,
   \label{two-inq}
   \end{eqnarray}
indicating that $\lim_{\Gamma\rightarrow\infty}F_v\{\theta_q^{\mathrm{min}}\}$ is as close as possible to the exact lower bound, $\lim_{\Gamma\rightarrow\infty}F$, of real free energy $F$. 

\vspace{4pt}
To summarize, it has been shown by reformulating the Lieb--Narnhofer lower bound that our field-theoretic approach to the strongly coupled OCP has superiority in consistency. Further evaluating the next leading order in $1/\Gamma^{1/2}$ expansion (effectively $1/\Gamma$), we obtain the excess internal energy similar to Rosenfeld's one \cite{Rosen-inter} which interpolates between the Debye-H\"uckel bound (relevant in the weak coupling regime $\Gamma<<1$) \cite{Mermin} and the Lieb--Narnhofer bound for $\Gamma>>1$ \cite{Lieb}; the details will be presented elsewhere.

%




\vspace{10pt}
\hspace{-23pt}We acknowledge the financial support from the Ministry of Education, Science, Culture, and Sports of Japan.


\end{document}